\documentclass{iau} 
\usepackage{graphicx}
\usepackage{hyperref,natbib}

\title[INTEGRAL results on magnetars] 
{INTEGRAL contributions to magnetars and multimessenger astrophysics}

\author[Sandro Mereghetti]   
{Sandro Mereghetti$^1$}

\affiliation{$^1$INAF, IASF-Milano, v. A. Corti 12, I-20133 Milano, Italy \\ email: {\tt sandro.mereghetti@inaf.it} }

\pubyear{2022}
\volume{363}  
\jname{Neutron Star Astrophysics at the Crossroads: Magnetars and the Multimessenger Revolution}
\editors{E. Troja \& M. Baring, eds.}

\def\sgr {SGR 1806--20}
\def\qui {1E 1547.0--5408}
\def\dic {SGR J1935+2154}
\def\frb {FRB 20200120E}
\def\gle{GLEAMX J162759.5--523504.3}

\def\apjl{ApJ}
\def\apj{ApJ}                 
\def\apjl{ApJ}                
\def\apss{Ap\&SS}             
\def\aap{A\&A}                
\def\nat{Nature}              

\def\nar{New~Astr.~Rev.} 

\begin{document}

\maketitle

\begin{abstract}
The INTEGRAL satellite, in orbit since October 2002,  has significantly contributed to the study of magnetars and, thanks to its unique capabilities for the study of transient gamma-ray phenomena,  it is now  playing  an important role in multimessenger astrophysics.
The most recent results include the discovery of a peculiar burst from \dic , which gave the first observational evidence for the connection between magnetars and fast radio bursts, and extensive searches for bursting activity in peculiar sources, such as the repeating \frb\ in M81 and the ultra-long period magnetar candidate \gle .

\end{abstract}

\firstsection 
\section{Introduction}

The European Space Agency INTEGRAL satellite, launched in October 2002 and still operational \citep{kuu21},  is  devoted to observations in the hard  X-ray / soft $\gamma$-ray range  with high spectral and angular resolution.  
Its two main instruments, IBIS and SPI, provide high sensitivity with good imaging and spectroscopic capabilities  over a wide field of view ($\sim$900 deg$^2$) in the range $\sim$20 keV -- 10 MeV, and can also detect transient  $\gamma$-ray signals from every direction in the sky thanks to their active anti-coincidence shields.  This is particularly useful because    the highly eccentric orbit of INTEGRAL, with a period of 2.7 days, allows uninterrupted observations  for 85\% of the time (i.e. when the satellite is above the  radiation belts) of virtually the whole sky 
(the fraction of sky occulted by the Earth goes from 0.05\% at perigee to 0.4\% when the satellite  enters/exits the radiation belts).  All the data are  downlinked in real time and processed
with a latency of only a few seconds from the time of their on-board acquisition allowing rapid localization of transient events.  
These unique properties  make INTEGRAL  a  powerful tool for time-domain and multimessenger astrophysics. A thorough description of the performances for the search of GW counterparts and other transient events can be found in \citet{sav17}. In this short contribution I concentrate on the results obtained for magnetars and fast radio bursts (FRB). See \citet{pap20} and \citet{fer21} for more extended accounts.

\section{Discovery of persistent hard X-ray emission from magnetars}
\label{sec:HX}

Magnetars were discovered in hard X-rays, but for many years their detection in this band was limited to the bursts from soft gamma-ray repeaters (SGR). 
Before the advent of  INTEGRAL, their persistent emission was seen  only below 10 keV,  with very soft spectra suggesting undetectable fluxes at higher energies.    
The unanticipated  discovery of persistent hard X-ray emission from several magnetars with INTEGRAL \citep{mol04,mer05a,kui06,goe06b,den08} opened a new window in the study of these objects.

The hard spectral components  in magnetars extend up to $\sim$150--200 keV with  power-law spectra ($\Gamma$$\sim$0.5-2) and contain an energy similar to, or even larger than,  that of the soft X-ray emission.  The pulsed flux is generally harder than the unpulsed one, implying an increase of pulsed fraction with energy.  
Contrary to the soft X-rays, most likely originating from thermal emission from the star surface, hard X-rays are produced by non-thermal magnetospheric particles, with an important contribution from resonant cyclotron scattering \citep{bel13,bar07}.

\begin{figure}[bht]
\begin{center}
 \includegraphics[width=9cm,angle=90]{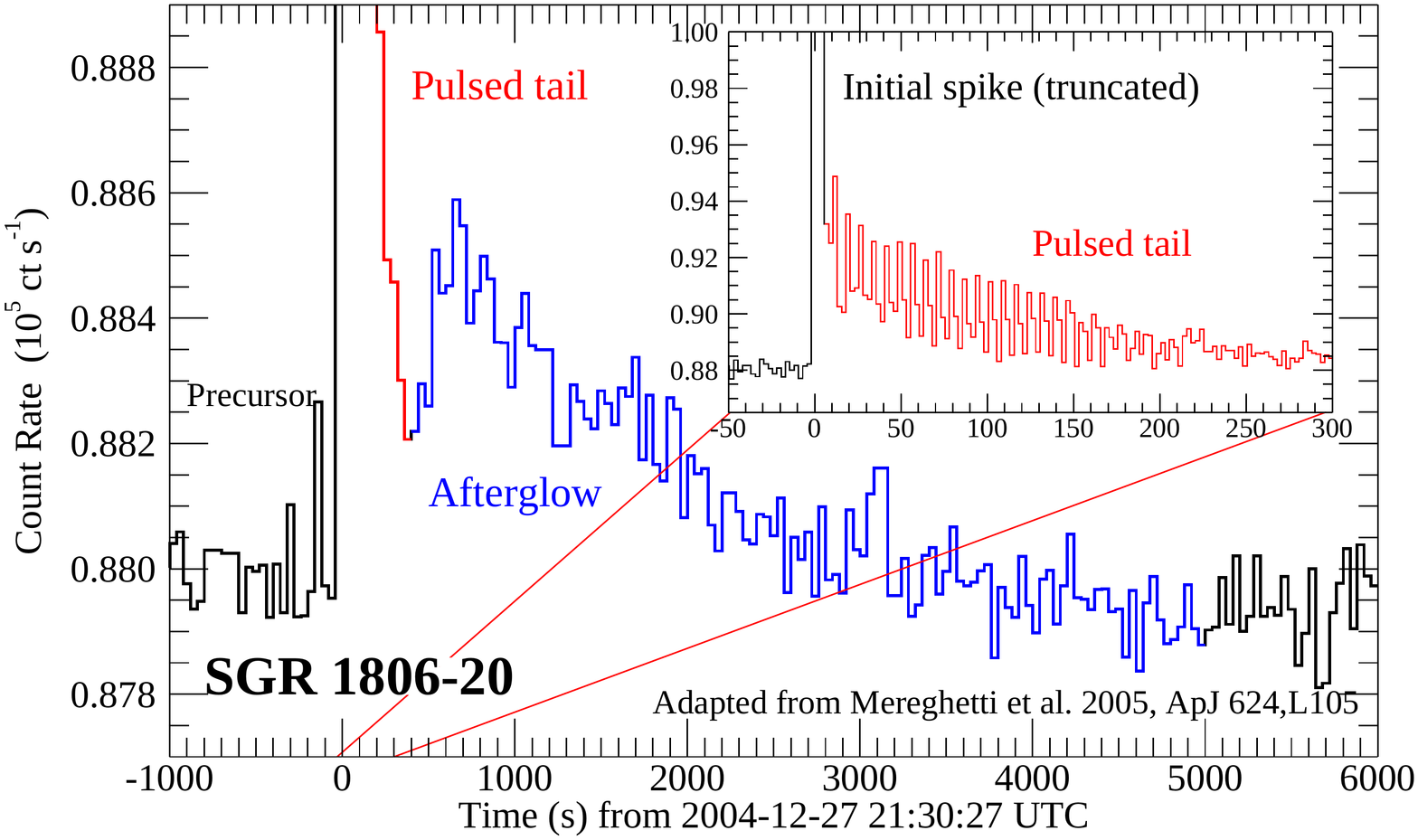} 
\caption{Light curve of the 2004 giant flare from \sgr .}
   \label{fig1}
\end{center}
\end{figure}

\section{The hard X-ray afterglow of the \sgr\ giant flare}

It is well known that magnetars giant flares (GF) consist of two main distinct components: an initial very bright and short ($<$0.2 s) spike with hard spectrum and a softer pulsed tail lasting a few minutes. 
Thanks to the sensitivity and uninterrupted coverage provided by the anti-coinicidence shield of the SPI instrument, INTEGRAL could discover a further distinct emission component in the  2004 December 27 GF from \sgr\ \citep{mer05b}.   
With a peak isotropic luminosity of $\sim10^{47}$ erg s$^{-1}$ and a  total energy release of $\sim10^{46}$ erg (for d=9 kpc), this was the most energetic GF ever seen from a Galactic magnetar.  
  
%
%
%
The light curve obtained with INTEGRAL (Fig.~\ref{fig1}) showed that, after the end of the pulsating tail, the hard X-ray flux above $\sim$80 keV increased again, reaching a peak  $\sim$700 s after the start of the GF, and   returning to the pre-flare background level after $\sim$4000 s.  
The time decay of such ``afterglow'' component was well fit by a power law  with $F(t)\propto  t^{-0.85}$. 
The fluence   in the 400-4000 s time interval was  of the same order of that in the pulsating tail.
The power-law time evolution and  the hard power law spectrum  ($\Gamma\sim$1.6, \citealt{fre07}) indicate that  this long-lasting emission is most likely caused by the interaction of relativistic ejecta   with the circumstellar material,  similar to the afterglows seen in    $\gamma$-ray bursts.  With standard  synchrotron models for  $\gamma$-ray burst afterglows, it is possible to relate the bulk Lorentz factor  of the ejected material, $\gamma_{ej}$, to  the time $t_{0}$ of the
afterglow onset. This gives
$\gamma_{ej}\sim$15($E$/5$\times$10$^{43}$ erg)$^{1/8}$($n$/0.1
cm$^{-3}$)$^{-1/8}$($t_{0}$/100)$^{-3/8}$, where $n$ is the
ambient density, consistent with a mildly
relativistic outflow, as it was also inferred from the analysis of the
radio source that appeared after the giant flare \citep{gra06}.

 
  
\section{Bursts from magnetars and fast radio bursts}

INTEGRAL also provided data on many short SGR bursts.    \sgr\ \citep{goe04,goe06b}
and \qui\ \citep{mer09,sav10} were oberved during periods of intense bursting activity and,  thanks to the high sensitivity of the  IBIS instrument, it was possible to study the  properties of bursts down to fluences as faint  as $2\times10^{-8}$ erg cm$^{-2}$, 
Many  bursts were localized in real time by the INTEGRAL Burst Alert System (IBAS, \citealt{mer03}), when they occurred in the IBIS field of view. 

\begin{figure}[bht]
\begin{center}
 \includegraphics[width=14cm]{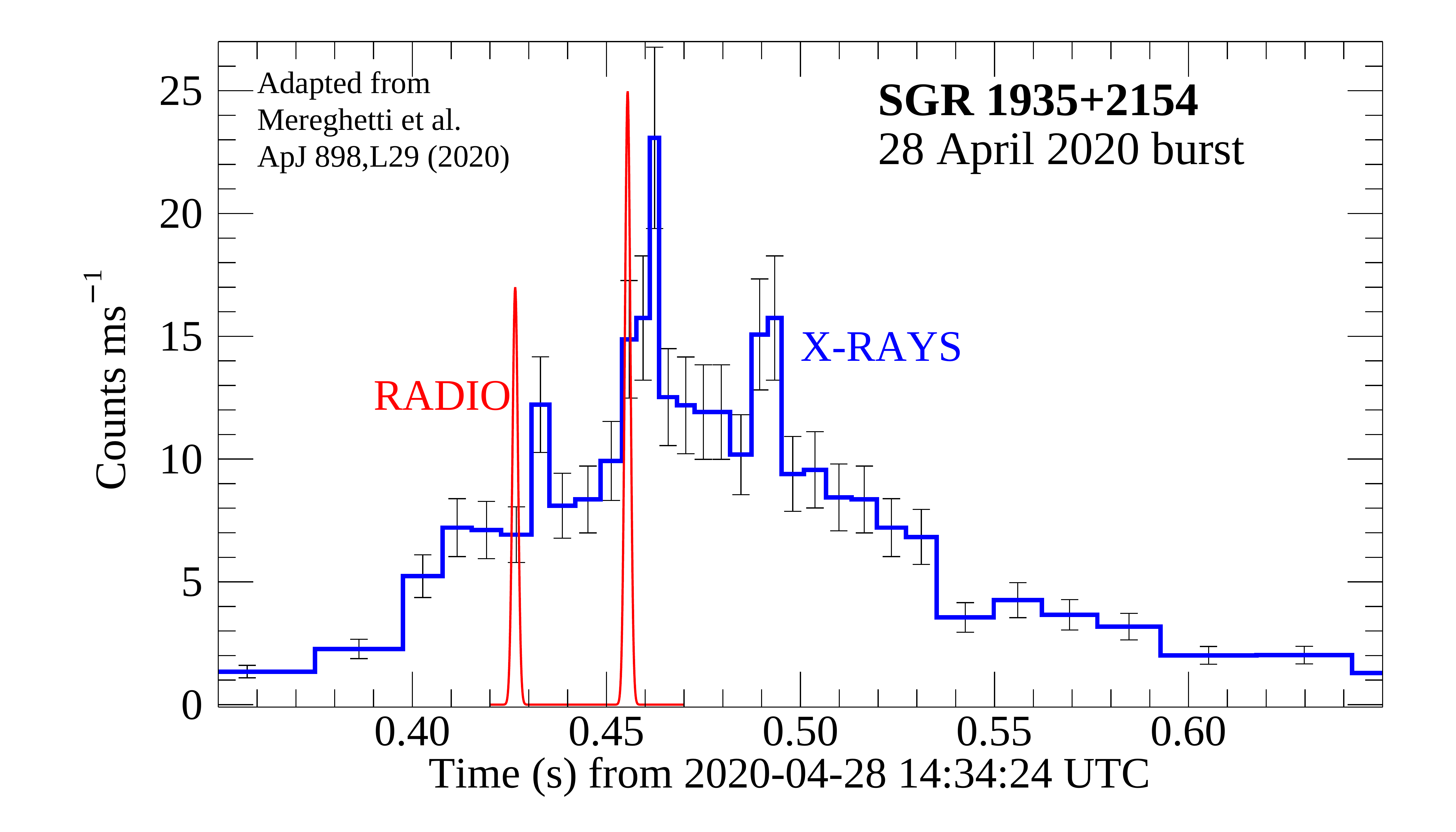} 
\caption{Light cuve of the 2020 April 28 burst from \dic .}
   \label{fig2}
\end{center}
\end{figure}

A particularly exciting case was that of the 2020 April 28 burst from \dic . This event triggered IBAS  \citep{mer20atel}, which distributed an alert after less than 10 s from the burst,  a few hours before the announcement of the independent discovery of the associated radio emission. The  properties of this very bright and short radio burst were very similar to those of the FRBs \citep{CHI20,boc20}, thus giving a strong support to FRB models involving magnetars (see \citet{zha20} and ref. therein). 
In hard X-rays this burst was not particularly  energetic  (E$\sim 10^{39}$ erg  for d=4.4 kpc, \citealt{mer20}), but it was characterized by a spectrum peaking at $\sim$70 keV, harder than that of the typical  bursts from this and other SGRs.
The  IBIS light curve (Fig.~\ref{fig2}) showed a broad bump, lasting $\sim$0.7 s with superimposed narrow spikes separated by $\sim$29 ms,  the same separation of  the two pulses seen at  400-800 MHz. It is thus tempting to associate the radio and X-ray pulses to the same phenomenon. Interestingly, the X-ray pulses  have a delay of 6.5 ms compared to the radio ones \citep{mer20}.

The  discovery  of simultaneous bursting emission at radio and high-energies from \dic\ gave a renewed impetus to  searches for high-energy emission associated to FRBs.  INTEGRAL took part in multiwavelength campaigns targeted at repeating FRBs  (e.g.  \citealt{pil20}), but the large distances of these extragalactic sources hampers their detection at X/$\gamma$-rays and the current limits are not very constraining. The number of relatively close FRBs is now increasing thanks to extensive observations with new radio facilities.  The repeating \frb\ has recently been discovered in  the  M81 galaxy   at a distance of only 3.6 Mpc  \citep{bha21}. 
INTEGRAL has repeatedly observed the region of M81, collecting a net exposure time of about 18 Ms. A search for hard X-ray bursts from \frb\ in these archival data gave negative results, but the derived limits are the best currently available  in the hard X-ray range for an extragalactic FRB \citep{mer21}. Although typical SGR-like bursts with 20-200 keV luminosity below $\sim10^{45} \left ( \frac{10~ms}{\Delta t} \right )$ erg s$^{-1}$ cannot be excluded ($\Delta t$  is the burst duration), these limits rule out the  emission of intermediate and giant flares from \frb .  Such events are quite rare in the known Galactic magnetars, but they might occur much more frequently in very young hyper-active magnetars that have been postulated in some FRB models. Note, however, that the location  in a globular cluster \citep{kir21} also suggests that \frb\ is not a young object, consistently with the INTEGRAL result, and might not be representative of the bulk FRB population.

\section{Long period magnetars}
\label{sec:lp}

Most magnetars have periods in the 1-12 s range, but there are a few peculiar sources showing signs of magnetar-like activity which have significantly longer periods. One of them is the central X-ray source in the RCW 103 supernova remnant, an apparently isolated neutron star with a period of 6.7 hr \citep{del06}. Another intriguing source recently discovered in the radio band, \gle,  has been reported at this conference. In 2018 January-March it emitted radio pulses, resembling those of magnetars,  with a periodicity of 1091 s \citep{hur22}. INTEGRAL observed its sky location for about 20 Ms from 2003 to 2021, but a search for magnetar-like hard X-ray bursts with the same procedure used for the  M81 FRB \citep{mer21}, did not reveal any activity from this putative magnetar.  For the reported distance of 1.3 kpc, the INTEGRAL limits rule out the emission of  bursts with 
L$\gtrsim2\times10^{38} \left ( \frac{10~ms}{\Delta t} \right )$ erg s$^{-1}$, well below the typical luminosity of SGR bursts.

  \vspace*{-.20 cm}
 

\end{document}